\documentclass[11pt,amssymb,nofootinbib]{revtex4}
\usepackage[latin9]{inputenc}
\usepackage{textcomp}
\usepackage{amsmath}
\usepackage{amssymb}
\usepackage{esint}
\usepackage{graphicx}
\usepackage{epstopdf}
\usepackage[english]{babel}

\usepackage[latin9]{inputenc}
\usepackage{textcomp}
\usepackage{amsmath}
\usepackage{amssymb}
\usepackage{esint}
\usepackage{graphicx}
\usepackage{epstopdf}
\usepackage[english]{babel}
\usepackage{verbatim}
\usepackage{bm}
\usepackage{color,graphicx}
\usepackage{tabularx}
\usepackage{times}

\begin{document}

\title{Seven non-standard models coupling quantum matter and gravity }
\author{Sandro Donadi}
\affiliation{Istituto Nazionale di Fisica Nucleare, Trieste Section, Via Valerio 2, 34127 Trieste, Italy}
\author{Angelo Bassi}
\affiliation{Department of Physics, University of Trieste, Strada Costiera 11, 34151 Trieste, Italy\looseness=-1}
\affiliation{Istituto Nazionale di Fisica Nucleare, Trieste Section, Via Valerio 2, 34127 Trieste, Italy.}

\begin{abstract}
    We review seven models, which consistently couple quantum matter and (Newtonian) gravity in a non standard way. For each of them we present the underlying motivations, the main equations and, when available, a comparison with experimental data. 
\end{abstract}

\maketitle

\section{Introduction}

Quantum Mechanics (QM) and General Relativity (GR) are the two pillars of modern physics; in their regime of validity, they work extremely well and their predictions have been confirmed with great accuracy.
One expects that there should be a unified theory that, in the appropriate regimes, reduce to QM and GR: to find such a theory is one of the biggest open challenges in theoretical physics. 

The most explored and in some sense obvious approach for merging the two theories is to quantize gravity. Essentially, given the classical variables of GR, i.e. the metric $g_{\mu\nu}$ and the stress-energy tensor $T_{\mu\nu}$ satisfying Einstein's equations
\begin{equation}\label{einstein}
    G_{\mu\nu}=\frac{8\pi G}{c^{4}}T_{\mu\nu},
\end{equation}
one notes that matter has been successfully quantized (modulo the interpretation issues of QM)  by promoting  the stress-energy tensor to the operator level: $T_{\mu\nu}\rightarrow \hat{T}_{\mu\nu}$; then for consistency also the metric tensor needs to be quantized: $G_{\mu\nu}\rightarrow \hat{G}_{\mu\nu}$. Several attempts have been explored, but a consistent quantum theory of gravity is still lacking \cite{rovelli2008loop,green2012superstring}. 

Here, we consider seven proposals which aim at combining QM and GR by focusing on the second logical  possibility: instead of directly quantizing space-time, one modifies QM in order to accomodate GR. In Penrose's words, the idea of these approaches is to achieve a ``gravitization  QM" \cite{penrose2014gravitization}, instead of quantizing GR. A  point in favor for this approach is the fact that modifications of quantum mechanics were also  suggested independently from gravity, to resolve the measurement problem in QM \cite{schrodinger1935current, leggett1980macroscopic, weinberg2014precision, bell2004speakable, ghirardi1986unified}. 

In the following we present each proposal, focusing on their motivations and assumptions, their main equations, point of strength and weaknesses and, when available, comparisons with experimental data. Specifically, in section \ref{kar} we review the proposal by Karolyhazy; in section \ref{secdiosis} we introduce  Di\'osi's model and in section \ref{penrose} the related proposal by Penrose; in section~\ref{Adler_sec} we review a model  proposed by Adler; in section \ref{SN_sec} we present the Schr\"odinger-Newton equation, while in  sections \ref{ktm} and \ref{tilloy} we discuss more recent proposals based on quantum measurement and feedback approaches.   

\section{The proposal by Karolyhazy }\label{kar}
One of the first attempts to combine QM and GR without quantizing the space-time metric is due to Karolyhazy \cite{karolyhazy1966gravitation,karolyhazy1982nuovo,k1986}. His main idea is that there is a fundamental uncertainty in the space-time structure, which can be modelled by considering the metric $g_{\mu\nu}$ as a random variable. To characterize this randomness,   Karolyhazy considers the following question: given a quantum probe subject to the uncertainty principle, and given a length $s=cT$, what is the minimal uncertainty $\Delta s$ with which it can be determined by the probe? 
He shows that $\Delta s$ can be estimated as \cite{karolyhazy1966gravitation}: 
\begin{equation}\label{deltas}
\Delta s\sim\sqrt[3]{\ell_{P}^{2}s},    
\end{equation}
where $\ell_{P}=\sqrt{\hbar G/c^{3}}$ is the Planck length. This uncertainty is then related to fluctuations of the space-time metric and consequently a random family of space time metrics $g_{\mu\nu}$ is introduced. 
To each realization of the metric, a length ``$s$" is associated, with  average $\Bar{s}=\mathbb{E}[s]$ and standard deviation $\overline{\Delta s}=\sqrt{\mathbb{E}[(s-\Bar{s})^2]}$, where $\mathbb{E}[...]$ denotes the average over the metric fluctuations.   

In the non-relativistic limit, the random deviations from flat space time affect only the component $g_{00}$ of the metric, hence one assumes $g_{00}=\eta_{00}+\gamma(\boldsymbol{x},t)$ and the fluctuations $\gamma(\boldsymbol{x},t)$ admit the Fourier expansion
\begin{equation}\label{espansione}
    \gamma(\boldsymbol{x},t)=\frac{1}{\sqrt{\ell^3}}\sum_{\boldsymbol{k}} \left(c(\boldsymbol{k})e^{i(\boldsymbol{k}\cdot\boldsymbol{x}-\omega t)}+c^*(\boldsymbol{k})e^{-i(\boldsymbol{k}\cdot\boldsymbol{x}-\omega t)}\right)
\end{equation}
where $\ell$ is the length of an arbitrarily chosen large box. In order to recover Eq. (\ref{deltas}) the correlations among the coefficients of the expansion must be:
\begin{equation}\label{corr_C}
    \mathbb{E}[c(\boldsymbol{k})c^{*}(\boldsymbol{k}')]=\begin{cases}
\delta_{\boldsymbol{k},\boldsymbol{k}'}\ell_{P}^{4/3}k^{-5/3} & \text{if}\; |k|\leq2\pi/\lambda_{c}\\
0 & \text{if}\; |k|>2\pi/\lambda_{c}
\end{cases}
\end{equation}
where $\lambda_c$ is a cutoff whose originally suggested value was $\lambda_c=10^{-15}$ m. The corresponding expression for the correlation among the perturbations $\gamma(\boldsymbol{x},t)$ was computed in \cite{bera2015comparison} for $\lambda_c=0$; the expression valid for any $\lambda_c$  is: 
\begin{align}\label{corr_Kar}
C(\boldsymbol{x}-\boldsymbol{x}',t-t'):=\mathbb{E}[\gamma(\boldsymbol{x},t)\gamma(\boldsymbol{x}',t)]&=\frac{3}{r}\left(\frac{\ell_{P}^{4}}{32\pi^{2}\lambda_{c}^{4}}\right)^{\frac{1}{3}}\left[(r+c\tau)\,_{1}F_{2}\left(\frac{2}{3};\frac{3}{2},\frac{5}{3};-\frac{\pi^{2}(r+c\tau)^{2}}{\lambda_{c}^{2}}\right)+\right.\nonumber\\
&\left.+(r-c\tau)\,_{1}F_{2}\left(\frac{2}{3};\frac{3}{2},\frac{5}{3};-\frac{\pi^{2}(r-c\tau)^{2}}{\lambda_{c}^{2}}\right)\right]
\end{align}
where  $r=|\boldsymbol{x}-\boldsymbol{x}'|$, $\tau=t-t'$ and $_q F_p (a;b;c;z)$ is the generalized hypergeometric function Gamma. 

The stochastic perturbation of the metric appears in the Schr\"odinger equation as a stochastic potential  of the form \cite{bera2015comparison} $V(\boldsymbol{x},t)=\frac{1}{2}m c^2 \gamma(\boldsymbol{x},t)$  (see also the discussion in section \ref{SN_sec} below). One can then perturbatively compute the corresponding master equation which is, to the lowest order in $G$: 
\begin{align}\label{gg}
    \frac{d\hat{\rho}(t)}{dt}&=-\frac{i}{\hbar}\left[\hat{H},\hat{\rho}(t)\right]-\left(\frac{c^{2}}{2\hbar}\right)^{2}\int d\boldsymbol{x}\int d\boldsymbol{x}'\int_{0}^{t}dt'C(\boldsymbol{x}-\boldsymbol{x}',t-t')\times\nonumber\\
    &\times\left[\varrho(\boldsymbol{x}-\hat{\boldsymbol{q}}),\left[e^{\frac{i}{\hbar}\hat{H}(t'-t)}\varrho(\boldsymbol{x}'-\hat{\boldsymbol{q}})e^{-\frac{i}{\hbar}\hat{H}(t'-t)},\hat{\rho}(t)\right]\right].
\end{align}
where $\varrho(\boldsymbol{x})$ is the mass density of the particle. In contrast to other models introduced below, this master equation is non-Markovian. 

The uncertainty in the space-time metric implies decoherence in position, as it should be clear from Eq. (\ref{gg}). Karolyhazy shows that for a uniform sphere of radius $R$, the randomness in space-time makes a quantum state decohere in position with a decay time $\tau_K$  corresponding to
\[
    \tau_{K}\approx\frac{ma_{K}^{2}}{\hbar},\;\;\;\;\;\textrm{with}\;\;\;\;\;a_{K}=\begin{cases}
\hbar^{2}/Gm^{3} & \text{if}\; R\ll \hbar^{2}/Gm^{3}\\
\left(\hbar^{2}R^2/G m^3\right)^{1/3}& \text{if}\; R\gg \hbar^{2}/Gm^{3}
\end{cases}
\]
To show some significant examples of the model's predictions, one finds that for a proton $a_K=10^{23}$ m, corresponding to $\tau_K=10^{53}$ s, while for a macroscopic sphere with radius $R=1$ cm, $a_K=10^{-18}$ m, corresponding to $\tau_K=10^{-4}$ s.

Stochastic modifications of the Schr\"odinger equation of the kind considered here are well known to lead to emission of X-rays from charged particles \cite{fu1997spontaneous,adler2007photon,adler2013spontaneous,donadi2014emission}, as a consequence of the fluctuations in the particle's motion. Di\'osi and Luck\'as computed this effect for the Karolyhazy's model, showing that the emitted radiation is unreasonably large and hence ruling out the model \cite{diosi1993calculation}. In their analysis, Di\'osi and Luck\'as considered a large range of  values for the cutoff $\lambda_c$, upon which the emission depends, and showed, using several arguments, that any value $\lambda_c\geq10^{-15}$ m is excluded. On the other hand, smaller values for the cutoff were considered as meaningless by Karolyhazy himself~\cite{karolyhazy1966gravitation} (likely because taking smaller cutoff implies that the model resolves distances smaller than the proton radius, which might be not compatible with the assumption of working in the non-relativistic regime, although a clear statement from the author in this respect is lacking). Further work is needed to assess whether Karolyhazy's idea can still survive or is really incompatible with experimental evidence.

\section{The Di\'osi model}\label{secdiosis}

What is known in the literature under the name of ``Di\'osi-Penrose" (DP) model actually consists of two related but independent proposals. Therefore we introduced them separately, starting from the earlier  model  by Di\'osi.

A master equation describing gravitational decoherence was first given in \cite{diosi1987universal}. The starting point is similar to that of Karolyhazy and it was presented in a previous paper by Di\'osi and Luck\'as \cite{diosi1987favor}. First, it is pointed out that in any realistic measurement one never measures the gravitational field $\boldsymbol{g}(\boldsymbol{r},t)$ at the precise space-time point $(\boldsymbol{r},t)$ with infinite precision, rather one measures  $\tilde{\boldsymbol{g}}(\boldsymbol{r},t)=\frac{1}{VT}\int_{T}dt'\int_{V}d\boldsymbol{r}'\boldsymbol{g}(\boldsymbol{x}',t')$ averaged over a space-time volume $VT$ around that point, where $V$ is a space volume and $T$ a time interval. Then they showed that if the measurement is performed using a quantum probe, the Heisenberg uncertainty principle implies that the minimum precision for determining $\tilde{\boldsymbol{g}}$ is given by $\delta\tilde{\boldsymbol{g}}\approx G \hbar/VT$. This result can be modelled by assuming that  $\boldsymbol{g}(\boldsymbol{r},t)=\boldsymbol{g}_c(\boldsymbol{r},t)+\boldsymbol{g}_s(\boldsymbol{r},t)$, where $\boldsymbol{g}_c$ is the classical field and $\boldsymbol{g}_s$ a stochastic vector fluctuation where the components $g_s^j$ have zero average and correlation $\mathbb{E}[g^i_s(\boldsymbol{r},t)g^j_s(\boldsymbol{r}',t)]=\delta^{ij}G\hbar\delta(t-t')\delta(\boldsymbol{r}-\boldsymbol{r}')$. Since $\boldsymbol{g}=-\nabla \phi$, one gets:
\begin{equation}\label{corr_Diosi}
    \mathbb{E}[\phi(\boldsymbol{r},t)\phi(\boldsymbol{r}',t)]=\delta(t-t')\frac{G}{\hbar|\boldsymbol{r}-\boldsymbol{r}'|}.
\end{equation}
It is interesting to compare this expression with that of Karolyazy in Eq. (\ref{corr_Kar}). By introducing $\phi$ as a stochastic potential in the Schr\"odinger equation, similarly to what suggested by Karolyhazy, Di\'osi finds that the corresponding master equation is
\begin{equation}\label{masterD}
    \frac{d\hat{\rho}(t)}{dt}=-\frac{i}{\hbar}\left[\hat{H},\hat{\rho}(t)\right]-\frac{G}{2\hbar}\int d\boldsymbol{r}\int d\boldsymbol{r}'\frac{1}{|\boldsymbol{r}-\boldsymbol{r}'|}\left[\hat{\varrho}(\boldsymbol{r}),\left[\hat{\varrho}(\boldsymbol{r}'),\hat{\rho}_{t}\right]\right]
\end{equation}
where $\hat{\varrho}(\boldsymbol{y})=\sum_{j=1}^{N}\varrho_{j}(\boldsymbol{y}-\boldsymbol{\hat{x}}_j)$ with $\varrho_{j}(\boldsymbol{y})$ the mass density of the $j$-th nucleon of the system\footnote{Due to their smaller mass, the  contribution from the electrons can be neglect, compared to that of the atomic nuclei. Here nucleons are taken as fundamental constituent because the model is formulated in the non relativistic regime, therefore the energies are not high enough to resolve their structure.}.

Given this master equation, later Di\'osi  built a non-linear equation which describes the collapse of the wave function~\cite{diosi1989models}:
\begin{align}\label{diosi}
    d|\psi(t)\rangle&=\left[-\frac{i}{\hbar}\hat{H}dt+\int d\boldsymbol{x}\left(\hat{\varrho}(\boldsymbol{r})-\langle\hat{\varrho}(\boldsymbol{r})\rangle_{t}\right)dW_{t}(\boldsymbol{r})+\right.\\
    &\left.-\frac{G}{2\hbar}\int d\boldsymbol{r}\int d\boldsymbol{r}'\frac{\left(\hat{\varrho}(\boldsymbol{r})-\langle\hat{\varrho}(\boldsymbol{r})\rangle_{t}\right)\left(\hat{\varrho}(\boldsymbol{r}')-\langle\hat{\varrho}(\boldsymbol{r}')\rangle_{t}\right)}{|\boldsymbol{r}-\boldsymbol{r}'|}\right]|\psi(t)\rangle.\nonumber
\end{align}

The master equation (\ref{masterD}) implies decoherence in the position basis. The decoherence time can be estimated by neglecting the free evolution and solving Eq. (\ref{masterD})  in the position basis; the result is:
\begin{equation}
\langle\boldsymbol{x}|\hat{\rho}(t)|\boldsymbol{y}\rangle=\langle\boldsymbol{x}|\hat{\rho}(0)|\boldsymbol{y}\rangle e^{-t/\tau_{D}},
\end{equation}
where:
\begin{equation}\label{tau}
    \tau_{D}^{-1}=\frac{G}{2\hbar}\int d\boldsymbol{r}\int d\boldsymbol{r}'\frac{[\varrho(\boldsymbol{r}-\boldsymbol{x})-\varrho(\boldsymbol{r}-\boldsymbol{y})][\varrho(\boldsymbol{r}'-\boldsymbol{x})-\varrho(\boldsymbol{r}'-\boldsymbol{y})]}{|\boldsymbol{r}-\boldsymbol{r}'|}.
\end{equation}
If one takes a point-like mass density, the term on the RHS  diverges, due to the functional form of the Newtonian potential. For this reason, one should consider mass distributions which are finite, extended over a region of radius $R_0$. For nucleons, a natural choice would be $R_0=10^{-15}$ m, as originally suggested in \cite{diosi1987universal}. However, Eq. (\ref{masterD})  predicts that the kinetic energy of a system increases linearly in time, due to the noise, and it was shown that such a small value for $R_0$ leads to an unacceptable energy increase~\cite{ghirardi1990continuous}. 

A possible solution to this problem is to introduce  dissipative terms, which bound the energy; this was considered in \cite{bahrami2014role} and the resulting  model depends on two parameters: $R_0$ and the temperature $T$ at which systems thermalize; the standard DP model is recovered in the limit of infinite temperature. However, for $R_0=10^{-15}$ m and taking a   temperature  $T\approx 1$ K for the noise (comparable to that of the cosmic microwave background radiation), the model is not consistent for systems with mass smaller than $m=10^{11}$ amu, since it predicts too strong dissipative effects, which are experimentally excluded. To summarize, contrary to other collapse models \cite{smirne2015dissipative}, the dissipative generalization of Di\'osi's model is not straightforward. For this reason, to date, the  model  still depends only on the parameter $R_0$, which is taken as a free parameter larger than $10^{-15}$ m. 

Several  bounds on $R_0$ have been computed, by considering different experimental data. In all of them, the key element is that the noise underlying the model generates a diffusion of the (center of mass of) the system, which can be detected, or bounded, by experiments. The model dictates that the smaller $R_0$ the larger the effect \cite{ghirardi1990continuous}, therefore experimental data place a lower bound on the parameter.  Gravitational wave detectors give $R_0\geq 4\times10^{-14}$ m \cite{helou2017lisa}, the analysis of the power radiated by neutron stars gives $R_0\gtrsim 10^{-13}$ m \cite{tilloy2019neutron} and searches for spontaneous photon  emission from Germanium detectors give $R_0\geq 0.54\times10^{-10}$ m \cite{donadi2021underground}. Moreover, the recent analysis performed in \cite{vinante2021gravity} 
shows that from the power radiated by Neptune one can derive $R_0\gtrsim 3.7 \times 10^{-12}$ m, while data about the residual heat leak experiments performed in ultralow temperature cryostats give $R_0\gtrsim 4.6\times 10^{-12}$ m.  

\section{Penrose's proposal}\label{penrose}
The proposal by Penrose is based on the general idea that, instead of working on quantizing gravity, we should try to ``gravitaze" QM \cite{penrose1996gravity,penrose2014gravitization}. This means that QM needs to be modified when gravitational effects are large enough. Penrose  proposed a way to compute what is meant by ``large": he did so by providing two different arguments, one based on an uncertainty in the definition of the time-translation operator \cite{penrose1996gravity} and the other based on possible violations of the equivalence principle \cite{penrose2014gravitization}. 

In both cases, the idea (expressed in a non-relativistic language) is that a mass in a spatial superposition around two different locations $a$ and $b$ of space (at some fixed time), generates a superposition of space-time metrics, such that a test mass used to probe the gravitational field at a point $\boldsymbol{r}$ would experience a superposition of two different accelerations, $\boldsymbol{g}_{a}(\boldsymbol{r})$ and $\boldsymbol{g}_{b}(\boldsymbol{r})$, corresponding to the two different branches of superposition. The square of the difference between these two accelerations divided by $G$ is taken as a measure of the energy uncertainty due to the space-time superpositions at the point $\boldsymbol{r}$ and then this uncertainty is integrated over  space. This implies: 
\begin{equation}\label{deltaE}
    \Delta E =\frac{1}{G}\int d\boldsymbol{r}\,\left(\boldsymbol{g}_{a}(\boldsymbol{r})-\boldsymbol{g}_{b}(\boldsymbol{r})\right)^{2}=4\pi G\int d\boldsymbol{r}\,\int d\boldsymbol{r}'\frac{\left(\varrho_{a}(\boldsymbol{r}')-\varrho_{b}(\boldsymbol{r}')\right)\left(\varrho_{a}(\boldsymbol{r})-\varrho_{b}(\boldsymbol{r})\right)}{|\boldsymbol{r}-\boldsymbol{r}'|},
\end{equation}
where to find the final expression one uses the standard relations $\boldsymbol{g}(\boldsymbol{r})=-\nabla\phi(\boldsymbol{r})$ together with the Poisson equation $\nabla^{2}\phi(\boldsymbol{r})=4\pi G\varrho(\boldsymbol{r})$ and its solution.

Given $\Delta E$, Penrose suggests that, similarly to what happens in nuclear decay, one can apply the Heisenberg time-energy uncertainty principle to derive the decay time of the superposition: $\tau=\hbar/\Delta E$. One can immediately see by comparing Eq. (\ref{deltaE}) with Eq. (\ref{tau}) that, apart for an unimportant  factor $8\pi$, this decay time is the same as that predicted by Di\'osi; this is the reason why the two proposal are referred together as ``Di\'osi-Penrose".

Contrary to the approaches of Karolyhazy and D\'iosi, Penrose does not provide a dynamical equation for the state vector. The beauty and strength of his argument is to  rely solely on basic and general principles of QM and GR and from this to show how they lead to a conflict. Remarkably, contrary to what one might have originally expected, Penrose's argument shows that the friction between QM and GR might be seen at scales much less extreme than the Planck scale (the Planck time is  $\sim 10^{-44}$ s and the Planck length $\sim 10^{-35}$ m). For example, if one consider an homogeneous sphere with mass $m=10^{-12}$ Kg and radius $R=5\;\mu$m placed in a spatial superposition at a distance $d\gg R$, one gets $\tau=10^{-6}$ s. A more precise calculation accounts for the fact that matter is not uniformly distributed but it is concentrated around atomic nuclei~\cite{penrose2014gravitization}. Then for the same mass $m=10^{-12}$ Kg the gravity induced collapse results to be stronger, leading to $\tau\simeq 10^{-2}-10^{-3}$ s.

As direct approaches to test Penrose's idea, it has been proposed to study the motion of a crystal with a mass $\sim 10^{-12}$ Kg inside an optical cavity \cite{marshall2003towards} or to perform interference experiments with a Bose-Einstein condensate made of $10^9$ atoms \cite{howl2019exploring}. Other indirect (non-interferometric) approaches were discussed at the end of the previous section.

\section{Adler's proposal}\label{Adler_sec}

The model of Karolyhazy, as well as the first model proposed by D\'iosi, are both based on the idea that the spacetime metric has random fluctuations. These fluctuations, in Karolyhazy's model \cite{karolyhazy1966gravitation},  as well as in the first version of the model proposed by D\'iosi \cite{diosi1987universal}, are supposed to be real. Real fluctuations keep the dynamics linear though random and induce decoherence, not the collapse of the wave function; an anti-Hermitian coupling leading to a non-linear (and stochastic) dynamics, by imposing state-vector normalization, is necessary to achieve objective collapse \cite{adler2004quantum}. 

Starting from these premises, Adler~\cite{bell2016quantum} provides specific arguments as to why it is reasonable to assume the existence of a small fluctuating complex part in the classical metric (for example, complex-valued effective metrics appear in theories of modified gravity~\cite{krasnov2015gr}). These fluctuations are coupled to the quantum mass density operator and, in the weak field and non-relativistic limit, one obtains the following approximate Hamiltonian: 
\begin{align}\label{eqstategen}
  \hat{H}=\hat{H}_{0}+\xi c^2\int d\boldsymbol{x}\hat{\varrho}(\boldsymbol{x})w(\boldsymbol{x},t);
\end{align}
here $w$ is a complex valued noise with zero average and correlation $\mathbb{E}[w(\boldsymbol{x},t),w(\boldsymbol{y},t')]=D(\boldsymbol{x}-\boldsymbol{y},t-t')$, where $D$ is a generic complex function of order 1 and the only assumption, which is made, is the correlation to be space and time translational invariant. The positive parameter $\xi$ controls the strength of the noise. $\hat{\varrho}(\boldsymbol{x})$ is the non-relativistic mass density operator. 

Since the fluctuations are complex, the Hamiltonian is not Hermitian, the norm of the vector is not conserved, and by forcing norm conservation appropriate non-linear terms have to added.  
The corresponding master equation was computed in \cite{gasbarri2017gravity} to the lowest perturbation order in $\xi$:
\begin{align}\label{eqmegen}
   \frac{d\hat{\rho}_{t}}{dt} &= -\frac{i}{\hbar}\left[\hat{H}_{0},\hat{\rho}_{t}\right]-\frac{\xi^{2}c^{4}}{\hbar^{2}}\int d\boldsymbol{x}\int d\boldsymbol{y}\int_{0}^{t}d\tau D^R(\boldsymbol{x}-\boldsymbol{y},t-\tau)\left[\hat{\varrho}(\boldsymbol{x}),\left[\hat{\varrho}(\boldsymbol{y},\tau-t),\hat{\rho}_{t}\right]\right]\nonumber\\
   &-\frac{i\xi^{2}c^{4}}{\hbar^{2}}\int d\boldsymbol{x}\int d\boldsymbol{y}\int_{0}^{t}d\tau D^I(\boldsymbol{x}-\boldsymbol{y},t-\tau)\left[\hat{\varrho}(\boldsymbol{x}),\left\{\hat{\varrho}(\boldsymbol{y},\tau-t),\hat{\rho}_{t}\right\}\right],
\end{align}
where $D^R$ and $D^I$ are, respectively, the real and imaginary parts of the correlation function $D$, $\hat{\varrho}(\boldsymbol{y},\tau-t)$ is the mass density operator evolved in the interaction picture to the time $\tau-t$. Since the noise correlator is arbitrary, apart from the symmetries specified before, Eq.~(\ref{eqmegen}) represents a general class of models which includes and generalizes also those of Karolyhazy and D\'iosi; moreover, it is suitable for describing dissipative as well as non-Markovian effects. 

It was shown~\cite{gasbarri2017gravity} that Eq.~\eqref{eqmegen} fulfills the two most important properties required for a consistent collapse models: the non-unitary terms induce the collapse of the wave function  in space and, when composite systems are considered, their centre of mass collapses with an amplified rate, roughly proportional to the total mass of the system. 
If one restricts the model to a noise correlator which is purely real, Gaussian in space and  delta-correlated in time, Eq.~\eqref{eqmegen} formally reduces to the master equation of the Continuous Spontaneous Localization (CSL) model \cite{ghirardi1990markov}, and hence all the bounds set for  CSL can be mapped to this model \cite{carlesso2019collapse,gasbarri2017gravity} (see \cite{carlesso2022} for a recent review on the subject). 
\section{The Schr\"odinger-Newton equation}\label{SN_sec}

The Schr\"odinger-Newton (SN) equation was first introduced in \cite{diosi1984gravitation} and it is strongly related to semiclassical gravity. In semiclassical gravity \cite{moller1962theories,rosenfeld1963quantization}, the Einstein equations (\ref{einstein}) of GR  are modified by quantizing matter, therefore promoting the energy-momentum tensor to an operator, and then by taking its expectation value:
\begin{equation}\label{semiclassical}
    G_{\mu\nu}=\frac{8\pi G}{c^{4}}\langle\psi|\hat{T}_{\mu\nu}|\psi\rangle
\end{equation}
where $|\psi\rangle$ describes the state of the quantum matter. Gravity remains classical. 

In the weak field limit, one can expand the metric around flat spacetime: $g_{\mu\nu}=\eta_{\mu\nu}+h_{\mu\nu}$ with $|h_{\mu\nu}|\ll|\eta_{\mu\nu}|$; in the non-relativistic limit, the only relevant component of the metric is $g_{00}$. Then Eq. (\ref{semiclassical}) becomes \cite{bahrami2014schrodinger}:
\begin{equation}\label{semiclassical2}
    \nabla^2 h_{00}=-\frac{8\pi G}{c^{4}}\langle\psi|\hat{T}_{00}|\psi\rangle.
\end{equation}
In  linearized gravity, the interaction between gravity and matter is described by the Hamiltonian $\hat{H}_I=-\frac{1}{2}\int d \boldsymbol{x} h_{\mu\nu} \hat{T}^{\mu\nu}$; considering only the relevant ``$00$" component, and given that $\hat{T}^{00}=\hat{\varrho}c^2$ with $\hat{\varrho}=\sum_j m_j \hat{\psi}_j^\dagger\hat{\psi}_j$ (the sum runs over the different kinds of particles composing the system, and $\hat{\psi}_j^\dagger\hat{\psi}_j$ is the number density operator for the particles of type $j$) and for the solution of the Poisson equation (\ref{semiclassical2}) one gets:
\begin{equation}\label{Hint}
    \hat{H}_I=-G\int d\boldsymbol{x}\int d\boldsymbol{x}'\frac{\langle\psi|\hat{\varrho}(\boldsymbol{x}')|\psi\rangle}{|\boldsymbol{x}-\boldsymbol{x}'|}\hat{\varrho}(\boldsymbol{x}).
\end{equation}
When this potential is added into the Schr\"odinger equation, one obtains the SN equation, which for one particle is:     
\begin{equation}\label{SN}
    i\hbar\frac{d}{dt}\psi_{t}(\boldsymbol{x},t)=\left(-\frac{\hbar^{2}}{2m}\nabla^{2}-Gm^{2}\int d\boldsymbol{y}\frac{|\psi_{t}(\boldsymbol{y},t)|^{2}}{|\boldsymbol{x}-\boldsymbol{y}|}\right)\psi_{t}(\boldsymbol{x},t).
\end{equation} 
The second term on the right hand side describes a Newtonian gravitational self-attraction among the different parts of the wave function; for a free particle, this term prevents the spread of the wave packet to increase indefinitely, as the kinetic term dictates; eventually the two effects compensate and the wave function reaches asymptotically a final width.  The effect is negligible for microscopic systems over typical timescales, and becomes more relevant for large masses. For a Gaussian wave packet a simple way to estimate the size of the equilibrium width $\sigma$ is given by $\sigma\sim\frac{\hbar^{2}}{Gm^{3}}$ \cite{carlip2008quantum}, which for a proton is $\sim10^{22}$ m while for a particle with mass of 1 mg is $\sim10^{-40}$ m.  

The SN equation is non-linear and deterministic. This poses a serious problem because it was shown that this kind of dynamics allows for  faster than light signalling \cite{gisin1989stochastic}. In the case of the SN equation, this has been explicitly shown  in \cite{bahrami2014schrodinger} and the argument is the following: one considers the typical EPR setup involving an entangled state of pairs of massive spin 1/2 particles, one sent to Alice and the other one to Bob, who are arbitrarily far away from each other. Depending on which measurement Alice decides to perform, Bob will either have 50\% times $|\uparrow\rangle$ and 50\% times $|\downarrow\rangle$ along a given $z$ direction, or  50\% times $\frac{|\uparrow\rangle+|\downarrow\rangle}{\sqrt{2}}$ and 50\% times $\frac{|\uparrow\rangle-|\downarrow\rangle}{\sqrt{2}}$.  Bob then sends his particles through a Stern-Gerlach device, with the inhomogeneous component of the magnetic filed aligned along the $z$ direction. When exiting the device, in the first case particles' trajectories will simply be deviated upwards or downwards by the magnetic field. In the second case their trajectories will turn in  a superposition of moving  upward and  downwards; then, assuming that the masses are large enough so that  the self attraction of the SN equation is not negligible, the two parts of the wave function will attract each other. When hitting the screen producing two spots, in the first case the distance between the spots will be larger than in the second case. In this way, Bob can understand which type of measurement Alice performed, thus  
establishing a protocol for superluminal communication.

This poses serious issues in considering Schr\"odinger-Newton equation as a fundamental dynamics. Since the equation is  a special case of  semiclassical gravity, this implies also Eq. (\ref{semiclassical}) cannot be taken as fundamental dynamics. A possible approach to solve this problem is discussed in section \ref{tilloy}. 

Compared to the effects in the DP model, those of the SN equation are typically weaker, which is why the model has not yet been tested. However, possibilities for future tests has been analysed, in particular considering optomechanical devices, and it was shown that by using the best current technologies in the fields of cavity enhanced optical position readout, levitation of ions in Paul traps, and superconducting materials, one should be able to measure the shift in the energy levels of an harmonic oscillator due to the presence of the SN self-interaction~\cite{grossardt2016optomechanical}. Another proposal, based on the study of the effects of SN equation on the motion of the centre of mass of a Si crystal cooled down to 10 K, was considered in \cite{yang2013macroscopic}, where it was also found that the SN effects are small but in principle detectable with current technology. 

\section{The KTM model} \label{ktm}
The model proposed by Kafri, Taylor and Milburn (KTM) \cite{kafri2014classical} is based on the idea of implementing Newtonian gravity through a hybrid classical-quantum dynamics based on a continuous measurement and feedback protocol. 

The model considers two masses $m_1$ and $m_2$, each trapped by harmonic potentials with respectively frequencies $\omega_1$ and $\omega_2$, and interacting through the Newtonian potential. The distance $d$ between the masses is supposed to be much larger than the position spread of each particle, so that the Newtonian potential can be approximated by a linear potential, leading to the effective one dimensional
Hamiltonian:
\begin{equation}\label{H_KTM}
    \hat{H}= \hat{H}_0 + K\hat{x}_{1}\hat{x}_{2}, \qquad \hat{H}_0 = \sum_{j=1}^{2}\frac{\hat{p}_{j}}{2m_{j}}+\frac{1}{2}m_{j}\Omega_{j}^{2}\hat{x}_{j}^{2}
\end{equation}
where $\Omega_{j}^{2}=\omega_{j}^{2}-K/m_{j}$ and $K=2Gm_{1}m_{2}/d^{3}$.

The first input of the KTM model is that each particle is subject to a continuous measurement of their position~\cite{jacobs2006straightforward}; the measurement records $r_j(t)$ are random variables with average $\langle\hat{x}_{j}(t)\rangle$ and fluctuations proportional to white noises $w_{j}(t)$:
\begin{equation}\label{r(t)}
    r_{j}(t)=\langle\hat{x}_{j}(t)\rangle+\frac{\hbar}{\sqrt{\gamma_j}}w_{j}(t),
\end{equation}
where the parameters $\gamma_j$, which at this stage are arbitrary, control the information associated to the outcome of the continuous measurement. Next, the measurement records are used as a feedback to generate an hybrid classical-quantum (linearized) Newtonian interaction in place of the standard quantum interaction: 
\begin{equation}\label{modH}
K\hat{x}_{1}\hat{x}_{2}\;\;\longrightarrow\;\;K(r_{1}(t)\hat{x}_{2}+r_{2}(t)\hat{x}_{1});
\end{equation}
this is why one speaks of a continuous measurement and feedback protocol.

The continuous measurement introduces a disturbance on the evolution of the two particles, which eventually leads to decoherence. KTM show that in order to minimise the decoherence effects, one needs to set $\gamma_j=2\hbar K$, thus linking these constants to gravity. 

The evolution for the state vector is given by \cite{gaona2021gravitational}:
\begin{align}\label{KTM_psi}
    d|\psi(t)\rangle&\!=\!\left\{ \! - \frac{i}{\hbar} \hat{H}_0 dt  -\sum_{\underset{j\neq k}{j,k=1}}^{2}\left[\frac{i}{\hbar}K\hat{x}_{j}r_{k}(t)+\frac{K\hat{x}_{j}^{2}}{4\hbar}\right]dt +\right.\\
    &+\sum_{j=1}^{2}\!\left[-\frac{K}{4\hbar}(\hat{x}_{j}-\langle\hat{x}_{j}(t)\rangle)^{2}dt+\sqrt{\frac{K}{2\hbar}}(\hat{x}_{j}-\langle\hat{x}_{j}(t)\rangle)dW_{j}(t)\right]\nonumber\\
    &\left.-\sum_{\underset{j\neq k}{j,k=1}}^{2}\frac{i}{2\hbar}K\hat{x}_{j}(\hat{x}_{k}-\langle\hat{x}_{k}(t)\rangle)dt\right\} |\psi(t)\rangle, \nonumber
\end{align}
where the first term on the right-hand-side is the standard Hamiltonian term $\hat{H}_0$ of Eq.~\eqref{H_KTM} (without the gravitational part),  the second term corresponds to the feedback contribution, the third and fourth term to  the continuous measurement, while the last term is the It\^o term arising from their combined effect. As one can see, this is a stochastic and highly non-linear equation, where the gravitational interaction can be barely recognized. Yet, the corresponding master equation (when an average over the noise is taken) is 
\begin{equation}\label{KTM_rho}
    \frac{d\hat{\rho}(t)}{dt}=-\frac{i}{\hbar}\left[\hat{H}_0 +K\hat{x}_{1}\hat{x}_{2},\hat{\rho}(t)\right]-\frac{K}{2\hbar}\sum_{j=1}^{2}\left[\hat{x}_{j}\left[\hat{x}_{j},\hat{\rho}(t)\right]\right].
\end{equation}
This is a Lindblad equation, where now (linearized) gravity reappears in the unitary part of the dynamics as a standard quantum interaction. Therefore, while at the wave function level the system evolves through a hybrid classical-quantum dynamics, at the level of the master equation the dynamics is fully quantum, as the classical and nonlinear terms are averaged out. The price to pay is the appearance of a Lindblad term, describing decoherence in position.

The KTM model can be generalized to more than two particles in different ways, and a priori it is not obvious  which one is more appropriate. One generalization, introduced in \cite{altamirano2018gravity}, assumes a pairwise interaction: given a system of $N$ particles interacting gravitationally, the position of each particle is measured by all the others separately; therefore, the KTM protocol previously outlined is repeated for each pair. In the same paper, however, it is shown that such a generalization is ruled out by interferometric experiments with atomic fountains. Another natural generalization, introduced in \cite{gaona2021gravitational}, assumes that the position of each particle is measured only once, and then the recorded information is shared to all the other particles of the system through the KTM feedback Hamiltonian. However, in the same paper, this generalization is proved to be inconsistent because, given two composite systems and assuming the internal degrees of freedom of each system are negligible, by integrating them out one does not recover the KTM model for the centre of mass of the two subsystems. 

As a last remark, the KTM model  was recently generalized by introducing dissipative terms ~\cite{di2021gravity}. 

\section{The Tilloy-Di\'osi model}\label{tilloy}

The Tilloy-Di\'osi (TD) model \cite{tilloy2016sourcing} has several connections to the  models introduced in the previous sections: it was proposed as a solution of the faster-than-light signalling problem in semi-classical gravity discussed in section \ref{SN_sec}; it introduces a feedback mechanism similar to the one used in the KTM model presented in section~\ref{ktm}; in an appropriate limit, its master equation reduces to that of the DP model of section~\ref{secdiosis}.  

The problem with the nonlinear SN equation (\ref{SN}) is that it is  deterministic. To solve this problem, in a clever way Tilloy and Di\'osi  introduce random terms in  the dynamics, by assuming that the source of gravity $\langle\psi|\hat{\varrho}(\boldsymbol{r})|\psi\rangle$ is replaced by $\langle\psi|\hat{\varrho}(\boldsymbol{r})|\psi\rangle+\hbar\int d\boldsymbol{s}\gamma^{-1}(\boldsymbol{r}-\boldsymbol{s}) \delta\varrho_t(\boldsymbol{s})$, where $\gamma^{-1}$ in the inverse kernel of $\gamma$ in the sense that it satisfies $\int d\boldsymbol{s}\gamma(\boldsymbol{r}-\boldsymbol{s})\gamma^{-1}(\boldsymbol{s}-\boldsymbol{r})=\delta(\boldsymbol{r}-\boldsymbol{r}')$ and $\delta\varrho_t(\boldsymbol{s})$ is a stochastic fluctuation resulting from the continuous measurement of the mass density at the spacetime point $(t,\boldsymbol{s})$, having zero average and correlation $\mathbb{E}[\delta\varrho_{t}(\boldsymbol{s})\delta\varrho_{t'}(\boldsymbol{s}')]=\gamma(\boldsymbol{s}-\boldsymbol{s}')\delta(t-t')$. 

By resorting to the same continuous measurement and feedback mechanism employed in the KTM model, now generalzied to continuous measurements at each spacetime point, they arrive at a non-trivial stochastic non-linear evolution for the state vector $|\psi_t\rangle$ that generalizes that in Eq. (\ref{KTM_psi}) to the full Newtonian potential~\cite{gaona2021gravitational}. The average over the noise gives the Lindblad-type master equation:
\begin{equation}\label{ME_TD}
    \frac{d\hat{\rho}(t)}{dt}=-\frac{i}{\hbar}\left[\hat{H}+\hat{V}_{\text{\tiny NEW}},\hat{\rho}(t)\right]-\int d\boldsymbol{s}\int d\boldsymbol{r}D(\boldsymbol{r}-\boldsymbol{s})\left[\hat{\varrho}(\boldsymbol{s})\left[\hat{\varrho}(\boldsymbol{r}),\hat{\rho}(t)\right]\right],
\end{equation}
where $\hat{V}_{\text{\tiny NEW}}$ is the standard Newtonian potential and
\begin{equation}\label{D_td}
D(\boldsymbol{r}-\boldsymbol{s})=\frac{\gamma(\boldsymbol{r}-\boldsymbol{s})}{8\hbar^{2}}+\frac{G^{2}}{2}\int d\boldsymbol{r}'\int d\boldsymbol{s}'\frac{\gamma^{-1}(\boldsymbol{r}'-\boldsymbol{s}')}{|\boldsymbol{r}-\boldsymbol{r}'||\boldsymbol{s}-\boldsymbol{s}'|}
\end{equation}
is the decoherence kernel. Similarly to what done by KTM, one can ask for this kernel to be minimal, which fixes  $\gamma(\boldsymbol{r}-\boldsymbol{s})=\frac{2\hbar G}{|\boldsymbol{r}-\boldsymbol{s}|}$. With this choice, the Lindblad term in Eq. (\ref{ME_TD}) reduces to that of Eq.~\eqref{masterD} of the DP model. In spite of having the same decoherence term as that of the DP model, the TD model has the important merit of naturally accounting for the Newtonian  interaction, which is not accounted for in the DP model (it is introduced by hand in the Hamiltonian). 


Another approach that modifies the SN equation by adding stochastic elements was put forward in \cite{nimmrichter2015stochastic}. As for the TD model, the stochastic terms are built in such a way to guarantee no faster-than-light signalling; One difference with respect to the TD model is that using this approach one does not recover the Newtonian potential.

As a final note, it is important to point out that, contrary to what one would expect, the TD model does {\it not} reduce to the KTM in the limit where  gravity is linearized, as it was shown in \cite{gaona2021gravitational}. This is a consequence of the fact that the noises responsible for the measurements are structurally different in the two models: while in the KTM model  the noises are attached to the particles,  in the TD there is a noise at each point of space-time, independently from where the particles are located. This change of perspective is necessary in order to deal with the full gravitational potential, as discussed in~\cite{gaona2021gravitational}.

\section{Conclusions}

We reviewed some of the most relevant proposals, which attempt to combine quantum mechanics and gravity, by keeping gravity classical to some extent, and modifying quantum mechanics where necessary. In most of these approaches the idea is that the gravitational interaction coupling quantum matter is modified by adding a stochastic perturbation. Clearly  these models  share the limitation of being valid only in the non-relativistic Netwonian regime.  

Here we did not review {\it gravitational decoherence} \cite{Sanchez, power2000decoherence,breuer2009metric,blencowe2013effective,anastopoulos2013master,pikovski2015universal}, i.e. standard quantum mechanical models describing how gravitational perturbations can spoil the  properties of quantum systems, since the discussion covered only those proposals which modify the quantum-gravitational interaction at the fundamental level. For a recent review of this field, we refer the reader to \cite{bassi2017gravitational}; a general master equation comprising  most of these models has been recently presented~\cite{asprea2021gravitational}. 

As a final note, the question whether gravity is fundamentally quantum or classical has recently received a boost by a proposal, which show how to use optomechanical platforms to perform future dedicated experiments assessing the nature of gravity \cite{bose2017spin,marletto2017gravitationally}. 


\section*{Acknowledgements}

AB acknowledges financial support from the H2020 FET Project TEQ (Grant No. 766900), the CNR/RS (London) project ``Testing fundamental theories with ultracold atoms'', the Foundational Questions Institute and Fetzer Franklin Fund, a donor advised fund of Silicon Valley Community Foundation (Grant No. FQXi-RFP-CPW- 2002), INFN and the University of Trieste. SD acknowledges financial support from INFN.

\vspace{0.5cm}
\noindent{\bf Data availability statement}

\noindent{}Data sharing is not applicable to this article as no new data were created or analyzed in this study.

\vspace{0.5cm}
\noindent{\bf Author Declarations}

\noindent{}The authors have no conflicts to disclose.

\bibliography{Articoli.bib}{}

\begin{thebibliography}{62}%
\makeatletter
\providecommand \@ifxundefined [1]{%
 \@ifx{#1\undefined}
}%
\providecommand \@ifnum [1]{%
 \ifnum #1\expandafter \@firstoftwo
 \else \expandafter \@secondoftwo
 \fi
}%
\providecommand \@ifx [1]{%
 \ifx #1\expandafter \@firstoftwo
 \else \expandafter \@secondoftwo
 \fi
}%
\providecommand \natexlab [1]{#1}%
\providecommand \enquote  [1]{``#1''}%
\providecommand \bibnamefont  [1]{#1}%
\providecommand \bibfnamefont [1]{#1}%
\providecommand \citenamefont [1]{#1}%
\providecommand \href@noop [0]{\@secondoftwo}%
\providecommand \href [0]{\begingroup \@sanitize@url \@href}%
\providecommand \@href[1]{\@@startlink{#1}\@@href}%
\providecommand \@@href[1]{\endgroup#1\@@endlink}%
\providecommand \@sanitize@url [0]{\catcode `\\12\catcode `\$12\catcode
  `\&12\catcode `\#12\catcode `\^12\catcode `\_12\catcode `\%12\relax}%
\providecommand \@@startlink[1]{}%
\providecommand \@@endlink[0]{}%
\providecommand \url  [0]{\begingroup\@sanitize@url \@url }%
\providecommand \@url [1]{\endgroup\@href {#1}{\urlprefix }}%
\providecommand \urlprefix  [0]{URL }%
\providecommand \Eprint [0]{\href }%
\providecommand \doibase [0]{http://dx.doi.org/}%
\providecommand \selectlanguage [0]{\@gobble}%
\providecommand \bibinfo  [0]{\@secondoftwo}%
\providecommand \bibfield  [0]{\@secondoftwo}%
\providecommand \translation [1]{[#1]}%
\providecommand \BibitemOpen [0]{}%
\providecommand \bibitemStop [0]{}%
\providecommand \bibitemNoStop [0]{.\EOS\space}%
\providecommand \EOS [0]{\spacefactor3000\relax}%
\providecommand \BibitemShut  [1]{\csname bibitem#1\endcsname}%
\let\auto@bib@innerbib\@empty
\bibitem [{\citenamefont {Rovelli}(2008)}]{rovelli2008loop}%
  \BibitemOpen
  \bibfield  {author} {\bibinfo {author} {\bibfnamefont {C.}~\bibnamefont
  {Rovelli}},\ }\href@noop {} {\bibfield  {journal} {\bibinfo  {journal}
  {Living reviews in relativity}\ }\textbf {\bibinfo {volume} {11}},\ \bibinfo
  {pages} {1} (\bibinfo {year} {2008})}\BibitemShut {NoStop}%
\bibitem [{\citenamefont {Green}\ \emph {et~al.}(2012)\citenamefont {Green},
  \citenamefont {Schwarz},\ and\ \citenamefont
  {Witten}}]{green2012superstring}%
  \BibitemOpen
  \bibfield  {author} {\bibinfo {author} {\bibfnamefont {M.~B.}\ \bibnamefont
  {Green}}, \bibinfo {author} {\bibfnamefont {J.~H.}\ \bibnamefont {Schwarz}},
  \ and\ \bibinfo {author} {\bibfnamefont {E.}~\bibnamefont {Witten}},\
  }\href@noop {} {\emph {\bibinfo {title} {Superstring theory: volume 2, loop
  amplitudes, anomalies and phenomenology}}}\ (\bibinfo  {publisher} {Cambridge
  university press},\ \bibinfo {year} {2012})\BibitemShut {NoStop}%
\bibitem [{\citenamefont {Penrose}(2014)}]{penrose2014gravitization}%
  \BibitemOpen
  \bibfield  {author} {\bibinfo {author} {\bibfnamefont {R.}~\bibnamefont
  {Penrose}},\ }\href@noop {} {\bibfield  {journal} {\bibinfo  {journal}
  {Foundations of Physics}\ }\textbf {\bibinfo {volume} {44}},\ \bibinfo
  {pages} {557} (\bibinfo {year} {2014})}\BibitemShut {NoStop}%
\bibitem [{\citenamefont {Schr{\"o}dinger}(1935)}]{schrodinger1935current}%
  \BibitemOpen
  \bibfield  {author} {\bibinfo {author} {\bibfnamefont {E.}~\bibnamefont
  {Schr{\"o}dinger}},\ }\href@noop {} {\bibfield  {journal} {\bibinfo
  {journal} {Naturwissenschaften}\ }\textbf {\bibinfo {volume} {23}},\ \bibinfo
  {pages} {807} (\bibinfo {year} {1935})}\BibitemShut {NoStop}%
\bibitem [{\citenamefont {Leggett}(1980)}]{leggett1980macroscopic}%
  \BibitemOpen
  \bibfield  {author} {\bibinfo {author} {\bibfnamefont {A.~J.}\ \bibnamefont
  {Leggett}},\ }\href@noop {} {\bibfield  {journal} {\bibinfo  {journal}
  {Progress of Theoretical Physics Supplement}\ }\textbf {\bibinfo {volume}
  {69}},\ \bibinfo {pages} {80} (\bibinfo {year} {1980})}\BibitemShut {NoStop}%
\bibitem [{\citenamefont {Weinberg}(1989)}]{weinberg2014precision}%
  \BibitemOpen
  \bibfield  {author} {\bibinfo {author} {\bibfnamefont {S.}~\bibnamefont
  {Weinberg}},\ }\href@noop {} {\bibfield  {journal} {\bibinfo  {journal}
  {Physical Review Letters}\ }\textbf {\bibinfo {volume} {62}},\ \bibinfo
  {pages} {485} (\bibinfo {year} {1989})}\BibitemShut {NoStop}%
\bibitem [{\citenamefont {Bell}(2004)}]{bell2004speakable}%
  \BibitemOpen
  \bibfield  {author} {\bibinfo {author} {\bibfnamefont {J.~S.}\ \bibnamefont
  {Bell}},\ }\href@noop {} {\emph {\bibinfo {title} {Speakable and unspeakable
  in quantum mechanics: Collected papers on quantum philosophy}}}\ (\bibinfo
  {publisher} {Cambridge university press},\ \bibinfo {year}
  {2004})\BibitemShut {NoStop}%
\bibitem [{\citenamefont {Ghirardi}\ \emph {et~al.}(1986)\citenamefont
  {Ghirardi}, \citenamefont {Rimini},\ and\ \citenamefont
  {Weber}}]{ghirardi1986unified}%
  \BibitemOpen
  \bibfield  {author} {\bibinfo {author} {\bibfnamefont {G.~C.}\ \bibnamefont
  {Ghirardi}}, \bibinfo {author} {\bibfnamefont {A.}~\bibnamefont {Rimini}}, \
  and\ \bibinfo {author} {\bibfnamefont {T.}~\bibnamefont {Weber}},\
  }\href@noop {} {\bibfield  {journal} {\bibinfo  {journal} {Physical review
  D}\ }\textbf {\bibinfo {volume} {34}},\ \bibinfo {pages} {470} (\bibinfo
  {year} {1986})}\BibitemShut {NoStop}%
\bibitem [{\citenamefont {Karolyhazy}(1966)}]{karolyhazy1966gravitation}%
  \BibitemOpen
  \bibfield  {author} {\bibinfo {author} {\bibfnamefont {F.}~\bibnamefont
  {Karolyhazy}},\ }\href@noop {} {\bibfield  {journal} {\bibinfo  {journal} {Il
  Nuovo Cimento A (1965-1970)}\ }\textbf {\bibinfo {volume} {42}},\ \bibinfo
  {pages} {390} (\bibinfo {year} {1966})}\BibitemShut {NoStop}%
\bibitem [{\citenamefont {K{\'a}rolyh{\'a}zy}\ \emph
  {et~al.}(1982)\citenamefont {K{\'a}rolyh{\'a}zy}, \citenamefont {Frenkel},\
  and\ \citenamefont {Luk{\'a}cs}}]{karolyhazy1982nuovo}%
  \BibitemOpen
  \bibfield  {author} {\bibinfo {author} {\bibfnamefont {F.}~\bibnamefont
  {K{\'a}rolyh{\'a}zy}}, \bibinfo {author} {\bibfnamefont {A.}~\bibnamefont
  {Frenkel}}, \ and\ \bibinfo {author} {\bibfnamefont {B.}~\bibnamefont
  {Luk{\'a}cs}},\ }\href@noop {} {\emph {\bibinfo {title} {On the Possibility
  of Observing the Eventual Breakdown of the Superposition Principle}}}\
  (\bibinfo  {publisher} {in Physics as Natural Philosophy, pp 204, edited by
  A. Simony and H. Feschbach, MIT Press, Cambridge, MA},\ \bibinfo {year}
  {1982})\BibitemShut {NoStop}%
\bibitem [{\citenamefont {K{\'a}rolyh{\'a}zy}\ \emph
  {et~al.}(1986)\citenamefont {K{\'a}rolyh{\'a}zy}, \citenamefont {Frenkel},\
  and\ \citenamefont {Luk{\'a}cs}}]{k1986}%
  \BibitemOpen
  \bibfield  {author} {\bibinfo {author} {\bibfnamefont {F.}~\bibnamefont
  {K{\'a}rolyh{\'a}zy}}, \bibinfo {author} {\bibfnamefont {A.}~\bibnamefont
  {Frenkel}}, \ and\ \bibinfo {author} {\bibfnamefont {B.}~\bibnamefont
  {Luk{\'a}cs}},\ }\href@noop {} {\emph {\bibinfo {title} {On the possible role
  of gravity in the reduction of the wave function}}}\ (\bibinfo  {publisher}
  {in Quantum concepts in space and time, pp 109--128, edited by R. Penrose and
  C.J. Isham, Oxford University Press},\ \bibinfo {year} {1986})\BibitemShut
  {NoStop}%
\bibitem [{\citenamefont {Bera}\ \emph {et~al.}(2015)\citenamefont {Bera},
  \citenamefont {Donadi}, \citenamefont {Lochan},\ and\ \citenamefont
  {Singh}}]{bera2015comparison}%
  \BibitemOpen
  \bibfield  {author} {\bibinfo {author} {\bibfnamefont {S.}~\bibnamefont
  {Bera}}, \bibinfo {author} {\bibfnamefont {S.}~\bibnamefont {Donadi}},
  \bibinfo {author} {\bibfnamefont {K.}~\bibnamefont {Lochan}}, \ and\ \bibinfo
  {author} {\bibfnamefont {T.~P.}\ \bibnamefont {Singh}},\ }\href@noop {}
  {\bibfield  {journal} {\bibinfo  {journal} {Foundations of Physics}\ }\textbf
  {\bibinfo {volume} {45}},\ \bibinfo {pages} {1537} (\bibinfo {year}
  {2015})}\BibitemShut {NoStop}%
\bibitem [{\citenamefont {Fu}(1997)}]{fu1997spontaneous}%
  \BibitemOpen
  \bibfield  {author} {\bibinfo {author} {\bibfnamefont {Q.}~\bibnamefont
  {Fu}},\ }\href@noop {} {\bibfield  {journal} {\bibinfo  {journal} {Physical
  Review A}\ }\textbf {\bibinfo {volume} {56}},\ \bibinfo {pages} {1806}
  (\bibinfo {year} {1997})}\BibitemShut {NoStop}%
\bibitem [{\citenamefont {Adler}\ and\ \citenamefont
  {Ramazano{\u{g}}lu}()}]{adler2007photon}%
  \BibitemOpen
  \bibfield  {author} {\bibinfo {author} {\bibfnamefont {S.~L.}\ \bibnamefont
  {Adler}}\ and\ \bibinfo {author} {\bibfnamefont {F.~M.}\ \bibnamefont
  {Ramazano{\u{g}}lu}},\ }\href@noop {} {\bibfield  {journal} {\bibinfo
  {journal} {Journal of Physics A}\ }\textbf {\bibinfo {volume} {40}},\
  \bibinfo {pages} {13395 (2007); Journal of Physics A {\bf 42}, 109801
  (2009).}}\BibitemShut {Stop}%
\bibitem [{\citenamefont {Adler}\ \emph {et~al.}(2013)\citenamefont {Adler},
  \citenamefont {Bassi},\ and\ \citenamefont {Donadi}}]{adler2013spontaneous}%
  \BibitemOpen
  \bibfield  {author} {\bibinfo {author} {\bibfnamefont {S.~L.}\ \bibnamefont
  {Adler}}, \bibinfo {author} {\bibfnamefont {A.}~\bibnamefont {Bassi}}, \ and\
  \bibinfo {author} {\bibfnamefont {S.}~\bibnamefont {Donadi}},\ }\href@noop {}
  {\bibfield  {journal} {\bibinfo  {journal} {Journal of Physics A:
  Mathematical and Theoretical}\ }\textbf {\bibinfo {volume} {46}},\ \bibinfo
  {pages} {245304} (\bibinfo {year} {2013})}\BibitemShut {NoStop}%
\bibitem [{\citenamefont {Donadi}\ and\ \citenamefont
  {Bassi}(2014)}]{donadi2014emission}%
  \BibitemOpen
  \bibfield  {author} {\bibinfo {author} {\bibfnamefont {S.}~\bibnamefont
  {Donadi}}\ and\ \bibinfo {author} {\bibfnamefont {A.}~\bibnamefont {Bassi}},\
  }\href@noop {} {\bibfield  {journal} {\bibinfo  {journal} {Journal of Physics
  A: Mathematical and Theoretical}\ }\textbf {\bibinfo {volume} {48}},\
  \bibinfo {pages} {035305} (\bibinfo {year} {2014})}\BibitemShut {NoStop}%
\bibitem [{\citenamefont {Di{\'o}si}\ and\ \citenamefont
  {Luk{\'a}cs}(1993)}]{diosi1993calculation}%
  \BibitemOpen
  \bibfield  {author} {\bibinfo {author} {\bibfnamefont {L.}~\bibnamefont
  {Di{\'o}si}}\ and\ \bibinfo {author} {\bibfnamefont {B.}~\bibnamefont
  {Luk{\'a}cs}},\ }\href@noop {} {\bibfield  {journal} {\bibinfo  {journal}
  {Physics Letters A}\ }\textbf {\bibinfo {volume} {181}},\ \bibinfo {pages}
  {366} (\bibinfo {year} {1993})}\BibitemShut {NoStop}%
\bibitem [{\citenamefont {Diosi}(1987)}]{diosi1987universal}%
  \BibitemOpen
  \bibfield  {author} {\bibinfo {author} {\bibfnamefont {L.}~\bibnamefont
  {Diosi}},\ }\href@noop {} {\bibfield  {journal} {\bibinfo  {journal} {Physics
  letters A}\ }\textbf {\bibinfo {volume} {120}},\ \bibinfo {pages} {377}
  (\bibinfo {year} {1987})}\BibitemShut {NoStop}%
\bibitem [{\citenamefont {Di{\'o}si}\ and\ \citenamefont
  {Luk{\'a}cs}(1987)}]{diosi1987favor}%
  \BibitemOpen
  \bibfield  {author} {\bibinfo {author} {\bibfnamefont {L.}~\bibnamefont
  {Di{\'o}si}}\ and\ \bibinfo {author} {\bibfnamefont {B.}~\bibnamefont
  {Luk{\'a}cs}},\ }\href@noop {} {\bibfield  {journal} {\bibinfo  {journal}
  {Annalen der Physik}\ }\textbf {\bibinfo {volume} {499}},\ \bibinfo {pages}
  {488} (\bibinfo {year} {1987})}\BibitemShut {NoStop}%
\bibitem [{\citenamefont {Di{\'o}si}(1989)}]{diosi1989models}%
  \BibitemOpen
  \bibfield  {author} {\bibinfo {author} {\bibfnamefont {L.}~\bibnamefont
  {Di{\'o}si}},\ }\href@noop {} {\bibfield  {journal} {\bibinfo  {journal}
  {Physical Review A}\ }\textbf {\bibinfo {volume} {40}},\ \bibinfo {pages}
  {1165} (\bibinfo {year} {1989})}\BibitemShut {NoStop}%
\bibitem [{\citenamefont {Ghirardi}\ \emph
  {et~al.}(1990{\natexlab{a}})\citenamefont {Ghirardi}, \citenamefont
  {Grassi},\ and\ \citenamefont {Rimini}}]{ghirardi1990continuous}%
  \BibitemOpen
  \bibfield  {author} {\bibinfo {author} {\bibfnamefont {G.}~\bibnamefont
  {Ghirardi}}, \bibinfo {author} {\bibfnamefont {R.}~\bibnamefont {Grassi}}, \
  and\ \bibinfo {author} {\bibfnamefont {A.}~\bibnamefont {Rimini}},\
  }\href@noop {} {\bibfield  {journal} {\bibinfo  {journal} {Physical Review
  A}\ }\textbf {\bibinfo {volume} {42}},\ \bibinfo {pages} {1057} (\bibinfo
  {year} {1990}{\natexlab{a}})}\BibitemShut {NoStop}%
\bibitem [{\citenamefont {Bahrami}\ \emph
  {et~al.}(2014{\natexlab{a}})\citenamefont {Bahrami}, \citenamefont {Smirne},\
  and\ \citenamefont {Bassi}}]{bahrami2014role}%
  \BibitemOpen
  \bibfield  {author} {\bibinfo {author} {\bibfnamefont {M.}~\bibnamefont
  {Bahrami}}, \bibinfo {author} {\bibfnamefont {A.}~\bibnamefont {Smirne}}, \
  and\ \bibinfo {author} {\bibfnamefont {A.}~\bibnamefont {Bassi}},\
  }\href@noop {} {\bibfield  {journal} {\bibinfo  {journal} {Physical Review
  A}\ }\textbf {\bibinfo {volume} {90}},\ \bibinfo {pages} {062105} (\bibinfo
  {year} {2014}{\natexlab{a}})}\BibitemShut {NoStop}%
\bibitem [{\citenamefont {Smirne}\ and\ \citenamefont
  {Bassi}(2015)}]{smirne2015dissipative}%
  \BibitemOpen
  \bibfield  {author} {\bibinfo {author} {\bibfnamefont {A.}~\bibnamefont
  {Smirne}}\ and\ \bibinfo {author} {\bibfnamefont {A.}~\bibnamefont {Bassi}},\
  }\href@noop {} {\bibfield  {journal} {\bibinfo  {journal} {Scientific
  reports}\ }\textbf {\bibinfo {volume} {5}},\ \bibinfo {pages} {1} (\bibinfo
  {year} {2015})}\BibitemShut {NoStop}%
\bibitem [{\citenamefont {Helou}\ \emph {et~al.}(2017)\citenamefont {Helou},
  \citenamefont {Slagmolen}, \citenamefont {McClelland},\ and\ \citenamefont
  {Chen}}]{helou2017lisa}%
  \BibitemOpen
  \bibfield  {author} {\bibinfo {author} {\bibfnamefont {B.}~\bibnamefont
  {Helou}}, \bibinfo {author} {\bibfnamefont {B.}~\bibnamefont {Slagmolen}},
  \bibinfo {author} {\bibfnamefont {D.~E.}\ \bibnamefont {McClelland}}, \ and\
  \bibinfo {author} {\bibfnamefont {Y.}~\bibnamefont {Chen}},\ }\href@noop {}
  {\bibfield  {journal} {\bibinfo  {journal} {Physical Review D}\ }\textbf
  {\bibinfo {volume} {95}},\ \bibinfo {pages} {084054} (\bibinfo {year}
  {2017})}\BibitemShut {NoStop}%
\bibitem [{\citenamefont {Tilloy}\ and\ \citenamefont
  {Stace}(2019)}]{tilloy2019neutron}%
  \BibitemOpen
  \bibfield  {author} {\bibinfo {author} {\bibfnamefont {A.}~\bibnamefont
  {Tilloy}}\ and\ \bibinfo {author} {\bibfnamefont {T.~M.}\ \bibnamefont
  {Stace}},\ }\href@noop {} {\bibfield  {journal} {\bibinfo  {journal}
  {Physical review letters}\ }\textbf {\bibinfo {volume} {123}},\ \bibinfo
  {pages} {080402} (\bibinfo {year} {2019})}\BibitemShut {NoStop}%
\bibitem [{\citenamefont {Donadi}\ \emph {et~al.}(2021)\citenamefont {Donadi},
  \citenamefont {Piscicchia}, \citenamefont {Curceanu}, \citenamefont
  {Di{\'o}si}, \citenamefont {Laubenstein},\ and\ \citenamefont
  {Bassi}}]{donadi2021underground}%
  \BibitemOpen
  \bibfield  {author} {\bibinfo {author} {\bibfnamefont {S.}~\bibnamefont
  {Donadi}}, \bibinfo {author} {\bibfnamefont {K.}~\bibnamefont {Piscicchia}},
  \bibinfo {author} {\bibfnamefont {C.}~\bibnamefont {Curceanu}}, \bibinfo
  {author} {\bibfnamefont {L.}~\bibnamefont {Di{\'o}si}}, \bibinfo {author}
  {\bibfnamefont {M.}~\bibnamefont {Laubenstein}}, \ and\ \bibinfo {author}
  {\bibfnamefont {A.}~\bibnamefont {Bassi}},\ }\href@noop {} {\bibfield
  {journal} {\bibinfo  {journal} {Nature Physics}\ }\textbf {\bibinfo {volume}
  {17}},\ \bibinfo {pages} {74} (\bibinfo {year} {2021})}\BibitemShut {NoStop}%
\bibitem [{\citenamefont {Vinante}\ and\ \citenamefont
  {Ulbricht}(2021)}]{vinante2021gravity}%
  \BibitemOpen
  \bibfield  {author} {\bibinfo {author} {\bibfnamefont {A.}~\bibnamefont
  {Vinante}}\ and\ \bibinfo {author} {\bibfnamefont {H.}~\bibnamefont
  {Ulbricht}},\ }\href@noop {} {\bibfield  {journal} {\bibinfo  {journal} {AVS
  Quantum Science}\ }\textbf {\bibinfo {volume} {3}},\ \bibinfo {pages}
  {045602} (\bibinfo {year} {2021})}\BibitemShut {NoStop}%
\bibitem [{\citenamefont {Penrose}(1996)}]{penrose1996gravity}%
  \BibitemOpen
  \bibfield  {author} {\bibinfo {author} {\bibfnamefont {R.}~\bibnamefont
  {Penrose}},\ }\href@noop {} {\bibfield  {journal} {\bibinfo  {journal}
  {General relativity and gravitation}\ }\textbf {\bibinfo {volume} {28}},\
  \bibinfo {pages} {581} (\bibinfo {year} {1996})}\BibitemShut {NoStop}%
\bibitem [{\citenamefont {Marshall}\ \emph {et~al.}(2003)\citenamefont
  {Marshall}, \citenamefont {Simon}, \citenamefont {Penrose},\ and\
  \citenamefont {Bouwmeester}}]{marshall2003towards}%
  \BibitemOpen
  \bibfield  {author} {\bibinfo {author} {\bibfnamefont {W.}~\bibnamefont
  {Marshall}}, \bibinfo {author} {\bibfnamefont {C.}~\bibnamefont {Simon}},
  \bibinfo {author} {\bibfnamefont {R.}~\bibnamefont {Penrose}}, \ and\
  \bibinfo {author} {\bibfnamefont {D.}~\bibnamefont {Bouwmeester}},\
  }\href@noop {} {\bibfield  {journal} {\bibinfo  {journal} {Physical Review
  Letters}\ }\textbf {\bibinfo {volume} {91}},\ \bibinfo {pages} {130401}
  (\bibinfo {year} {2003})}\BibitemShut {NoStop}%
\bibitem [{\citenamefont {Howl}\ \emph {et~al.}(2019)\citenamefont {Howl},
  \citenamefont {Penrose},\ and\ \citenamefont {Fuentes}}]{howl2019exploring}%
  \BibitemOpen
  \bibfield  {author} {\bibinfo {author} {\bibfnamefont {R.}~\bibnamefont
  {Howl}}, \bibinfo {author} {\bibfnamefont {R.}~\bibnamefont {Penrose}}, \
  and\ \bibinfo {author} {\bibfnamefont {I.}~\bibnamefont {Fuentes}},\
  }\href@noop {} {\bibfield  {journal} {\bibinfo  {journal} {New Journal of
  Physics}\ }\textbf {\bibinfo {volume} {21}},\ \bibinfo {pages} {043047}
  (\bibinfo {year} {2019})}\BibitemShut {NoStop}%
\bibitem [{\citenamefont {Adler}(2004)}]{adler2004quantum}%
  \BibitemOpen
  \bibfield  {author} {\bibinfo {author} {\bibfnamefont {S.~L.}\ \bibnamefont
  {Adler}},\ }\href@noop {} {\emph {\bibinfo {title} {Quantum theory as an
  emergent phenomenon: The statistical mechanics of matrix models as the
  precursor of quantum field theory}}}\ (\bibinfo  {publisher} {Cambridge
  University Press},\ \bibinfo {year} {2004})\BibitemShut {NoStop}%
\bibitem [{\citenamefont {Adler}(2016)}]{bell2016quantum}%
  \BibitemOpen
  \bibfield  {author} {\bibinfo {author} {\bibfnamefont {S.}~\bibnamefont
  {Adler}},\ }\href@noop {} {\emph {\bibinfo {title} {Gravitation and the noise
  needed in objective reduction models}}}\ (\bibinfo  {publisher} {in Quantum
  Nonlocality and Reality: 50 Years of Bell's Theorem, ed Mary Bell and Shan
  Gao, Cambridge University Press},\ \bibinfo {year} {2016})\BibitemShut
  {NoStop}%
\bibitem [{\citenamefont {Krasnov}(2015)}]{krasnov2015gr}%
  \BibitemOpen
  \bibfield  {author} {\bibinfo {author} {\bibfnamefont {K.}~\bibnamefont
  {Krasnov}},\ }\href@noop {} {\bibfield  {journal} {\bibinfo  {journal}
  {Journal of High Energy Physics}\ }\textbf {\bibinfo {volume} {10}},\
  \bibinfo {pages} {paper 037} (\bibinfo {year} {2015})}\BibitemShut {NoStop}%
\bibitem [{\citenamefont {Gasbarri}\ \emph {et~al.}(2017)\citenamefont
  {Gasbarri}, \citenamefont {Toro{\v{s}}}, \citenamefont {Donadi},\ and\
  \citenamefont {Bassi}}]{gasbarri2017gravity}%
  \BibitemOpen
  \bibfield  {author} {\bibinfo {author} {\bibfnamefont {G.}~\bibnamefont
  {Gasbarri}}, \bibinfo {author} {\bibfnamefont {M.}~\bibnamefont
  {Toro{\v{s}}}}, \bibinfo {author} {\bibfnamefont {S.}~\bibnamefont {Donadi}},
  \ and\ \bibinfo {author} {\bibfnamefont {A.}~\bibnamefont {Bassi}},\
  }\href@noop {} {\bibfield  {journal} {\bibinfo  {journal} {Physical Review
  D}\ }\textbf {\bibinfo {volume} {96}},\ \bibinfo {pages} {104013} (\bibinfo
  {year} {2017})}\BibitemShut {NoStop}%
\bibitem [{\citenamefont {Ghirardi}\ \emph
  {et~al.}(1990{\natexlab{b}})\citenamefont {Ghirardi}, \citenamefont
  {Pearle},\ and\ \citenamefont {Rimini}}]{ghirardi1990markov}%
  \BibitemOpen
  \bibfield  {author} {\bibinfo {author} {\bibfnamefont {G.~C.}\ \bibnamefont
  {Ghirardi}}, \bibinfo {author} {\bibfnamefont {P.}~\bibnamefont {Pearle}}, \
  and\ \bibinfo {author} {\bibfnamefont {A.}~\bibnamefont {Rimini}},\
  }\href@noop {} {\bibfield  {journal} {\bibinfo  {journal} {Physical Review
  A}\ }\textbf {\bibinfo {volume} {42}},\ \bibinfo {pages} {78} (\bibinfo
  {year} {1990}{\natexlab{b}})}\BibitemShut {NoStop}%
\bibitem [{\citenamefont {Carlesso}\ and\ \citenamefont
  {Donadi}(2019)}]{carlesso2019collapse}%
  \BibitemOpen
  \bibfield  {author} {\bibinfo {author} {\bibfnamefont {M.}~\bibnamefont
  {Carlesso}}\ and\ \bibinfo {author} {\bibfnamefont {S.}~\bibnamefont
  {Donadi}},\ }\href@noop {} {\emph {\bibinfo {title} {Collapse models: main
  properties and the state of art of the experimental tests}}}\ (\bibinfo
  {publisher} {in Advances in Open Systems and Fundamental Tests of Quantum
  Mechanics, pp 1--13, edited by B. Vacchini, H.-P. Breuer and A. Bassi,
  Springer},\ \bibinfo {year} {2019})\BibitemShut {NoStop}%
\bibitem [{\citenamefont {Carlesso}\ \emph {et~al.}(2022)\citenamefont
  {Carlesso}, \citenamefont {Donadi}, \citenamefont {Ferialdi}, \citenamefont
  {Paternostro}, \citenamefont {Ulbricht},\ and\ \citenamefont
  {Bassi}}]{carlesso2022}%
  \BibitemOpen
  \bibfield  {author} {\bibinfo {author} {\bibfnamefont {M.}~\bibnamefont
  {Carlesso}}, \bibinfo {author} {\bibfnamefont {S.}~\bibnamefont {Donadi}},
  \bibinfo {author} {\bibfnamefont {L.}~\bibnamefont {Ferialdi}}, \bibinfo
  {author} {\bibfnamefont {M.}~\bibnamefont {Paternostro}}, \bibinfo {author}
  {\bibfnamefont {H.}~\bibnamefont {Ulbricht}}, \ and\ \bibinfo {author}
  {\bibfnamefont {A.}~\bibnamefont {Bassi}},\ }\href@noop {} {\bibfield
  {journal} {\bibinfo  {journal} {Nature Physics}\ }\textbf {\bibinfo {volume}
  {18}},\ \bibinfo {pages} {243} (\bibinfo {year} {2022})}\BibitemShut
  {NoStop}%
\bibitem [{\citenamefont {Di{\'o}si}(1984)}]{diosi1984gravitation}%
  \BibitemOpen
  \bibfield  {author} {\bibinfo {author} {\bibfnamefont {L.}~\bibnamefont
  {Di{\'o}si}},\ }\href@noop {} {\bibfield  {journal} {\bibinfo  {journal}
  {Physics Letters A}\ }\textbf {\bibinfo {volume} {105}},\ \bibinfo {pages}
  {199} (\bibinfo {year} {1984})}\BibitemShut {NoStop}%
\bibitem [{\citenamefont {M{\o}ller}\ \emph {et~al.}(1962)\citenamefont
  {M{\o}ller} \emph {et~al.}}]{moller1962theories}%
  \BibitemOpen
  \bibfield  {author} {\bibinfo {author} {\bibfnamefont {C.}~\bibnamefont
  {M{\o}ller}} \emph {et~al.},\ }\href@noop {} {\bibfield  {journal} {\bibinfo
  {journal} {Colloques Internationaux CNRS}\ }\textbf {\bibinfo {volume} {91}}
  (\bibinfo {year} {1962})}\BibitemShut {NoStop}%
\bibitem [{\citenamefont {Rosenfeld}(1963)}]{rosenfeld1963quantization}%
  \BibitemOpen
  \bibfield  {author} {\bibinfo {author} {\bibfnamefont {L.}~\bibnamefont
  {Rosenfeld}},\ }\href@noop {} {\bibfield  {journal} {\bibinfo  {journal}
  {Nuclear Physics}\ }\textbf {\bibinfo {volume} {40}},\ \bibinfo {pages} {353}
  (\bibinfo {year} {1963})}\BibitemShut {NoStop}%
\bibitem [{\citenamefont {Bahrami}\ \emph
  {et~al.}(2014{\natexlab{b}})\citenamefont {Bahrami}, \citenamefont
  {Gro{\ss}ardt}, \citenamefont {Donadi},\ and\ \citenamefont
  {Bassi}}]{bahrami2014schrodinger}%
  \BibitemOpen
  \bibfield  {author} {\bibinfo {author} {\bibfnamefont {M.}~\bibnamefont
  {Bahrami}}, \bibinfo {author} {\bibfnamefont {A.}~\bibnamefont
  {Gro{\ss}ardt}}, \bibinfo {author} {\bibfnamefont {S.}~\bibnamefont
  {Donadi}}, \ and\ \bibinfo {author} {\bibfnamefont {A.}~\bibnamefont
  {Bassi}},\ }\href@noop {} {\bibfield  {journal} {\bibinfo  {journal} {New
  Journal of Physics}\ }\textbf {\bibinfo {volume} {16}},\ \bibinfo {pages}
  {115007} (\bibinfo {year} {2014}{\natexlab{b}})}\BibitemShut {NoStop}%
\bibitem [{\citenamefont {Carlip}(2008)}]{carlip2008quantum}%
  \BibitemOpen
  \bibfield  {author} {\bibinfo {author} {\bibfnamefont {S.}~\bibnamefont
  {Carlip}},\ }\href@noop {} {\bibfield  {journal} {\bibinfo  {journal}
  {Classical and Quantum Gravity}\ }\textbf {\bibinfo {volume} {25}},\ \bibinfo
  {pages} {154010} (\bibinfo {year} {2008})}\BibitemShut {NoStop}%
\bibitem [{\citenamefont {Gisin}(1989)}]{gisin1989stochastic}%
  \BibitemOpen
  \bibfield  {author} {\bibinfo {author} {\bibfnamefont {N.}~\bibnamefont
  {Gisin}},\ }\href@noop {} {\bibfield  {journal} {\bibinfo  {journal} {Helv.
  Phys. Acta}\ }\textbf {\bibinfo {volume} {62}},\ \bibinfo {pages} {363}
  (\bibinfo {year} {1989})}\BibitemShut {NoStop}%
\bibitem [{\citenamefont {Gro{\ss}ardt}\ \emph {et~al.}(2016)\citenamefont
  {Gro{\ss}ardt}, \citenamefont {Bateman}, \citenamefont {Ulbricht},\ and\
  \citenamefont {Bassi}}]{grossardt2016optomechanical}%
  \BibitemOpen
  \bibfield  {author} {\bibinfo {author} {\bibfnamefont {A.}~\bibnamefont
  {Gro{\ss}ardt}}, \bibinfo {author} {\bibfnamefont {J.}~\bibnamefont
  {Bateman}}, \bibinfo {author} {\bibfnamefont {H.}~\bibnamefont {Ulbricht}}, \
  and\ \bibinfo {author} {\bibfnamefont {A.}~\bibnamefont {Bassi}},\
  }\href@noop {} {\bibfield  {journal} {\bibinfo  {journal} {Physical Review
  D}\ }\textbf {\bibinfo {volume} {93}},\ \bibinfo {pages} {096003} (\bibinfo
  {year} {2016})}\BibitemShut {NoStop}%
\bibitem [{\citenamefont {Yang}\ \emph {et~al.}(2013)\citenamefont {Yang},
  \citenamefont {Miao}, \citenamefont {Lee}, \citenamefont {Helou},\ and\
  \citenamefont {Chen}}]{yang2013macroscopic}%
  \BibitemOpen
  \bibfield  {author} {\bibinfo {author} {\bibfnamefont {H.}~\bibnamefont
  {Yang}}, \bibinfo {author} {\bibfnamefont {H.}~\bibnamefont {Miao}}, \bibinfo
  {author} {\bibfnamefont {D.-S.}\ \bibnamefont {Lee}}, \bibinfo {author}
  {\bibfnamefont {B.}~\bibnamefont {Helou}}, \ and\ \bibinfo {author}
  {\bibfnamefont {Y.}~\bibnamefont {Chen}},\ }\href@noop {} {\bibfield
  {journal} {\bibinfo  {journal} {Physical review letters}\ }\textbf {\bibinfo
  {volume} {110}},\ \bibinfo {pages} {170401} (\bibinfo {year}
  {2013})}\BibitemShut {NoStop}%
\bibitem [{\citenamefont {Kafri}\ \emph {et~al.}(2014)\citenamefont {Kafri},
  \citenamefont {Taylor},\ and\ \citenamefont {Milburn}}]{kafri2014classical}%
  \BibitemOpen
  \bibfield  {author} {\bibinfo {author} {\bibfnamefont {D.}~\bibnamefont
  {Kafri}}, \bibinfo {author} {\bibfnamefont {J.}~\bibnamefont {Taylor}}, \
  and\ \bibinfo {author} {\bibfnamefont {G.}~\bibnamefont {Milburn}},\
  }\href@noop {} {\bibfield  {journal} {\bibinfo  {journal} {New Journal of
  Physics}\ }\textbf {\bibinfo {volume} {16}},\ \bibinfo {pages} {065020}
  (\bibinfo {year} {2014})}\BibitemShut {NoStop}%
\bibitem [{\citenamefont {Jacobs}\ and\ \citenamefont
  {Steck}(2006)}]{jacobs2006straightforward}%
  \BibitemOpen
  \bibfield  {author} {\bibinfo {author} {\bibfnamefont {K.}~\bibnamefont
  {Jacobs}}\ and\ \bibinfo {author} {\bibfnamefont {D.~A.}\ \bibnamefont
  {Steck}},\ }\href@noop {} {\bibfield  {journal} {\bibinfo  {journal}
  {Contemporary Physics}\ }\textbf {\bibinfo {volume} {47}},\ \bibinfo {pages}
  {279} (\bibinfo {year} {2006})}\BibitemShut {NoStop}%
\bibitem [{\citenamefont {Gaona-Reyes}\ \emph {et~al.}(2021)\citenamefont
  {Gaona-Reyes}, \citenamefont {Carlesso},\ and\ \citenamefont
  {Bassi}}]{gaona2021gravitational}%
  \BibitemOpen
  \bibfield  {author} {\bibinfo {author} {\bibfnamefont {J.~L.}\ \bibnamefont
  {Gaona-Reyes}}, \bibinfo {author} {\bibfnamefont {M.}~\bibnamefont
  {Carlesso}}, \ and\ \bibinfo {author} {\bibfnamefont {A.}~\bibnamefont
  {Bassi}},\ }\href@noop {} {\bibfield  {journal} {\bibinfo  {journal}
  {Physical Review D}\ }\textbf {\bibinfo {volume} {103}},\ \bibinfo {pages}
  {056011} (\bibinfo {year} {2021})}\BibitemShut {NoStop}%
\bibitem [{\citenamefont {Altamirano}\ \emph {et~al.}(2018)\citenamefont
  {Altamirano}, \citenamefont {Corona-Ugalde}, \citenamefont {Mann},\ and\
  \citenamefont {Zych}}]{altamirano2018gravity}%
  \BibitemOpen
  \bibfield  {author} {\bibinfo {author} {\bibfnamefont {N.}~\bibnamefont
  {Altamirano}}, \bibinfo {author} {\bibfnamefont {P.}~\bibnamefont
  {Corona-Ugalde}}, \bibinfo {author} {\bibfnamefont {R.~B.}\ \bibnamefont
  {Mann}}, \ and\ \bibinfo {author} {\bibfnamefont {M.}~\bibnamefont {Zych}},\
  }\href@noop {} {\bibfield  {journal} {\bibinfo  {journal} {Classical and
  Quantum Gravity}\ }\textbf {\bibinfo {volume} {35}},\ \bibinfo {pages}
  {145005} (\bibinfo {year} {2018})}\BibitemShut {NoStop}%
\bibitem [{\citenamefont {Di~Bartolomeo}\ \emph {et~al.}(2021)\citenamefont
  {Di~Bartolomeo}, \citenamefont {Carlesso},\ and\ \citenamefont
  {Bassi}}]{di2021gravity}%
  \BibitemOpen
  \bibfield  {author} {\bibinfo {author} {\bibfnamefont {G.}~\bibnamefont
  {Di~Bartolomeo}}, \bibinfo {author} {\bibfnamefont {M.}~\bibnamefont
  {Carlesso}}, \ and\ \bibinfo {author} {\bibfnamefont {A.}~\bibnamefont
  {Bassi}},\ }\href@noop {} {\bibfield  {journal} {\bibinfo  {journal}
  {Physical Review D}\ }\textbf {\bibinfo {volume} {104}},\ \bibinfo {pages}
  {104027} (\bibinfo {year} {2021})}\BibitemShut {NoStop}%
\bibitem [{\citenamefont {Tilloy}\ and\ \citenamefont
  {Di{\'o}si}(2016)}]{tilloy2016sourcing}%
  \BibitemOpen
  \bibfield  {author} {\bibinfo {author} {\bibfnamefont {A.}~\bibnamefont
  {Tilloy}}\ and\ \bibinfo {author} {\bibfnamefont {L.}~\bibnamefont
  {Di{\'o}si}},\ }\href@noop {} {\bibfield  {journal} {\bibinfo  {journal}
  {Physical Review D}\ }\textbf {\bibinfo {volume} {93}},\ \bibinfo {pages}
  {024026} (\bibinfo {year} {2016})}\BibitemShut {NoStop}%
\bibitem [{\citenamefont {Nimmrichter}\ and\ \citenamefont
  {Hornberger}(2015)}]{nimmrichter2015stochastic}%
  \BibitemOpen
  \bibfield  {author} {\bibinfo {author} {\bibfnamefont {S.}~\bibnamefont
  {Nimmrichter}}\ and\ \bibinfo {author} {\bibfnamefont {K.}~\bibnamefont
  {Hornberger}},\ }\href@noop {} {\bibfield  {journal} {\bibinfo  {journal}
  {Physical Review D}\ }\textbf {\bibinfo {volume} {91}},\ \bibinfo {pages}
  {024016} (\bibinfo {year} {2015})}\BibitemShut {NoStop}%
\bibitem [{\citenamefont {Sanchez~Gomez}(1992)}]{Sanchez}%
  \BibitemOpen
  \bibfield  {author} {\bibinfo {author} {\bibfnamefont {J.}~\bibnamefont
  {Sanchez~Gomez}},\ }\href@noop {} {\bibfield  {journal} {\bibinfo  {journal}
  {Singapore: World Scientific}\ }\textbf {\bibinfo {volume} {456}},\ \bibinfo
  {pages} {88} (\bibinfo {year} {1992})}\BibitemShut {NoStop}%
\bibitem [{\citenamefont {Power}\ and\ \citenamefont
  {Percival}(2000)}]{power2000decoherence}%
  \BibitemOpen
  \bibfield  {author} {\bibinfo {author} {\bibfnamefont {W.}~\bibnamefont
  {Power}}\ and\ \bibinfo {author} {\bibfnamefont {I.}~\bibnamefont
  {Percival}},\ }\href@noop {} {\bibfield  {journal} {\bibinfo  {journal}
  {Proceedings of the Royal Society of London. Series A: Mathematical, Physical
  and Engineering Sciences}\ }\textbf {\bibinfo {volume} {456}},\ \bibinfo
  {pages} {955} (\bibinfo {year} {2000})}\BibitemShut {NoStop}%
\bibitem [{\citenamefont {Breuer}\ \emph {et~al.}(2009)\citenamefont {Breuer},
  \citenamefont {G{\"o}kl{\"u}},\ and\ \citenamefont
  {L{\"a}mmerzahl}}]{breuer2009metric}%
  \BibitemOpen
  \bibfield  {author} {\bibinfo {author} {\bibfnamefont {H.-P.}\ \bibnamefont
  {Breuer}}, \bibinfo {author} {\bibfnamefont {E.}~\bibnamefont
  {G{\"o}kl{\"u}}}, \ and\ \bibinfo {author} {\bibfnamefont {C.}~\bibnamefont
  {L{\"a}mmerzahl}},\ }\href@noop {} {\bibfield  {journal} {\bibinfo  {journal}
  {Classical and Quantum Gravity}\ }\textbf {\bibinfo {volume} {26}},\ \bibinfo
  {pages} {105012} (\bibinfo {year} {2009})}\BibitemShut {NoStop}%
\bibitem [{\citenamefont {Blencowe}(2013)}]{blencowe2013effective}%
  \BibitemOpen
  \bibfield  {author} {\bibinfo {author} {\bibfnamefont {M.}~\bibnamefont
  {Blencowe}},\ }\href@noop {} {\bibfield  {journal} {\bibinfo  {journal}
  {Physical review letters}\ }\textbf {\bibinfo {volume} {111}},\ \bibinfo
  {pages} {021302} (\bibinfo {year} {2013})}\BibitemShut {NoStop}%
\bibitem [{\citenamefont {Anastopoulos}\ and\ \citenamefont
  {Hu}(2013)}]{anastopoulos2013master}%
  \BibitemOpen
  \bibfield  {author} {\bibinfo {author} {\bibfnamefont {C.}~\bibnamefont
  {Anastopoulos}}\ and\ \bibinfo {author} {\bibfnamefont {B.}~\bibnamefont
  {Hu}},\ }\href@noop {} {\bibfield  {journal} {\bibinfo  {journal} {Classical
  and Quantum Gravity}\ }\textbf {\bibinfo {volume} {30}},\ \bibinfo {pages}
  {165007} (\bibinfo {year} {2013})}\BibitemShut {NoStop}%
\bibitem [{\citenamefont {Pikovski}\ \emph {et~al.}(2015)\citenamefont
  {Pikovski}, \citenamefont {Zych}, \citenamefont {Costa},\ and\ \citenamefont
  {Brukner}}]{pikovski2015universal}%
  \BibitemOpen
  \bibfield  {author} {\bibinfo {author} {\bibfnamefont {I.}~\bibnamefont
  {Pikovski}}, \bibinfo {author} {\bibfnamefont {M.}~\bibnamefont {Zych}},
  \bibinfo {author} {\bibfnamefont {F.}~\bibnamefont {Costa}}, \ and\ \bibinfo
  {author} {\bibfnamefont {{\v{C}}.}~\bibnamefont {Brukner}},\ }\href@noop {}
  {\bibfield  {journal} {\bibinfo  {journal} {Nature Physics}\ }\textbf
  {\bibinfo {volume} {11}},\ \bibinfo {pages} {668} (\bibinfo {year}
  {2015})}\BibitemShut {NoStop}%
\bibitem [{\citenamefont {Bassi}\ \emph {et~al.}(2017)\citenamefont {Bassi},
  \citenamefont {Gro{\ss}ardt},\ and\ \citenamefont
  {Ulbricht}}]{bassi2017gravitational}%
  \BibitemOpen
  \bibfield  {author} {\bibinfo {author} {\bibfnamefont {A.}~\bibnamefont
  {Bassi}}, \bibinfo {author} {\bibfnamefont {A.}~\bibnamefont {Gro{\ss}ardt}},
  \ and\ \bibinfo {author} {\bibfnamefont {H.}~\bibnamefont {Ulbricht}},\
  }\href@noop {} {\bibfield  {journal} {\bibinfo  {journal} {Classical and
  Quantum Gravity}\ }\textbf {\bibinfo {volume} {34}},\ \bibinfo {pages}
  {193002} (\bibinfo {year} {2017})}\BibitemShut {NoStop}%
\bibitem [{\citenamefont {Asprea}\ \emph {et~al.}(2021)\citenamefont {Asprea},
  \citenamefont {Gasbarri},\ and\ \citenamefont
  {Bassi}}]{asprea2021gravitational}%
  \BibitemOpen
  \bibfield  {author} {\bibinfo {author} {\bibfnamefont {L.}~\bibnamefont
  {Asprea}}, \bibinfo {author} {\bibfnamefont {G.}~\bibnamefont {Gasbarri}}, \
  and\ \bibinfo {author} {\bibfnamefont {A.}~\bibnamefont {Bassi}},\
  }\href@noop {} {\bibfield  {journal} {\bibinfo  {journal} {Physical Review
  D}\ }\textbf {\bibinfo {volume} {103}},\ \bibinfo {pages} {104041} (\bibinfo
  {year} {2021})}\BibitemShut {NoStop}%
\bibitem [{\citenamefont {Bose}\ \emph {et~al.}(2017)\citenamefont {Bose},
  \citenamefont {Mazumdar}, \citenamefont {Morley}, \citenamefont {Ulbricht},
  \citenamefont {Toro{\v{s}}}, \citenamefont {Paternostro}, \citenamefont
  {Geraci}, \citenamefont {Barker}, \citenamefont {Kim},\ and\ \citenamefont
  {Milburn}}]{bose2017spin}%
  \BibitemOpen
  \bibfield  {author} {\bibinfo {author} {\bibfnamefont {S.}~\bibnamefont
  {Bose}}, \bibinfo {author} {\bibfnamefont {A.}~\bibnamefont {Mazumdar}},
  \bibinfo {author} {\bibfnamefont {G.~W.}\ \bibnamefont {Morley}}, \bibinfo
  {author} {\bibfnamefont {H.}~\bibnamefont {Ulbricht}}, \bibinfo {author}
  {\bibfnamefont {M.}~\bibnamefont {Toro{\v{s}}}}, \bibinfo {author}
  {\bibfnamefont {M.}~\bibnamefont {Paternostro}}, \bibinfo {author}
  {\bibfnamefont {A.~A.}\ \bibnamefont {Geraci}}, \bibinfo {author}
  {\bibfnamefont {P.~F.}\ \bibnamefont {Barker}}, \bibinfo {author}
  {\bibfnamefont {M.}~\bibnamefont {Kim}}, \ and\ \bibinfo {author}
  {\bibfnamefont {G.}~\bibnamefont {Milburn}},\ }\href@noop {} {\bibfield
  {journal} {\bibinfo  {journal} {Physical review letters}\ }\textbf {\bibinfo
  {volume} {119}},\ \bibinfo {pages} {240401} (\bibinfo {year}
  {2017})}\BibitemShut {NoStop}%
\bibitem [{\citenamefont {Marletto}\ and\ \citenamefont
  {Vedral}(2017)}]{marletto2017gravitationally}%
  \BibitemOpen
  \bibfield  {author} {\bibinfo {author} {\bibfnamefont {C.}~\bibnamefont
  {Marletto}}\ and\ \bibinfo {author} {\bibfnamefont {V.}~\bibnamefont
  {Vedral}},\ }\href@noop {} {\bibfield  {journal} {\bibinfo  {journal}
  {Physical review letters}\ }\textbf {\bibinfo {volume} {119}},\ \bibinfo
  {pages} {240402} (\bibinfo {year} {2017})}\BibitemShut {NoStop}%
\end{thebibliography}%
\bibliographystyle{apsrev4-1}

\end{document}